\documentclass[aps,pre,floatfix,10pt]{revtex4}
\usepackage{graphicx}
\usepackage{dcolumn}
\usepackage{bm}
\usepackage{latexsym}
\usepackage{amsmath}
\usepackage{amssymb}
\usepackage{color}

\begin{document}
\title{Physics of collective transport and traffic phenomena in biology:\\
progress in 20 years}
\author{Debashish Chowdhury{\footnote{Corresponding Author; E-mail: debchg@gmail.com}}} 
\affiliation{Department of Physics, DIT University, Mussoorie Diversion Road, Dehradun 248009, India} 
\author{Andreas Schadschneider} 
\affiliation{Institute for Theoretical Physics, University of Cologne, 50937 K\"oln, Germany} 
\author{Katsuhiro Nishinari} 
\affiliation{Research Center for Advanced Science and Technology, University of Tokyo, Komaba 4-6-1, Meguro-ku, Tokyo 153-8904, Japan}
\date{\today}

\begin{abstract}
Enormous progress have been made in the last 20 years since the publication of our review \cite{csk05polrev} in this journal on transport and traffic phenomena in biology. In this brief article we present a glimpse of the major advances during this period. First, we present similarities and differences between collective intracellular transport of a single micron-size cargo by multiple molecular motors and that of a cargo particle by a team of ants on the basis of the common principle of load-sharing.  Second, we sketch several models all of which are biologically motivated extensions of the Asymmetric Simple Exclusion Process (ASEP); some of these models represent the traffic of molecular machines, like RNA polymerase (RNAP) and ribosome, that catalyze template-directed polymerization of RNA and proteins, respectively, whereas few other models capture the key features of the traffic of ants on trails. More specifically, using the ASEP-based models we demonstrate the effects of traffic of RNAPs and ribosomes on random and `programmed' errors in gene expression as well as on some other subcellular processes. We recall a puzzling empirical result on the single-lane traffic of predatory ants {\it Leptogenys processionalis} as well as recent attempts to account for this puzzle. We also mention some surprising effects of lane-changing rules observed in a ASEP-based model for 3-lane traffic of army ants. Finally, we explain the conceptual similarities between the pheromone-mediated indirect communication, called stigmergy, between ants on a trail and the floor-field-mediated interaction between humans in a pedestrian traffic. For the floor-field model of human pedestrian traffic we present a major theoretical result that is relevant from the perspective of all types of traffic phenomena.

\end{abstract} 

\maketitle

\tableofcontents{}

\section{Introduction: transport and traffic phenomena at different levels of organization in biology} 

It is quite common to hear ``it is moving! It is alive!''. Movement is often regarded as the hallmark of life. As is well known, a cell is the structural and functional unit of life. From the unicellular whale sperm to multicellular gigantic sperm whale, across seven orders of magnitude of length, movement is at the core of the biological processes that sustain life \cite{chong00}. Even the intracellular milieu of each living cell is like a bustling metropolis where various subcellular objects move on wide range of time scales. The laws of nature that govern the movement of inanimate matter have been the holy grail of the well established discipline of physics. The dynamical equations of physics describe how an object moves in space with the passage of time under the influence of external forces that can be derived from the respective fields (like gravitational, or electro-magnetic fields). However, the movements at the subcellular, cellular or organismal level are either cause or effects of biological functions. Therefore, in order to understand life, it is not enough to describe how a particular object moves but also to explain the functional role of these movements. What distinguishes the moving objects of our interest here from, for example, celestial bodies of Newtonian mechanics is that the latter are passive field-driven objects  whereas the former are active and self-driven in the sense that the force needed for their movement are generated by some internal chemical processes and not imposed from outside \cite{css00,scnbook}. Moreover, none of the moving objects is isolated from the others and the motion is not deterministic; therefore, a theoretical treatment of the large number of interacting moving objects needs to be formulated in the spirit of statistical mechanics. Furthermore, since living systems are far from equilibrium, appropriate adaptation of the formalisms of non-equilibrium statistical mechanics for passive field-driven objects to that for active self-driven objects is providing deep insight into physics of life \cite{css00,scnbook}.

Motor proteins move along specific filaments; their directed, albeit noisy, movements is driven by free energy extracted from a chemical reaction (called hydrolysis) catalyzed by the motor protein whereby a higher energy fuel molecule is converted to a lower energy product \cite{chowdhury13,kolomeiskybook}. A subcellular cargo (most common being vesicles and organelles), which is generally much bigger than a single motor protein, is transported by a team of motor proteins to a destination where the cargo is delivered for performing its own biological function \cite{mallik13,hancock14}. This collective transport of a micron-size cargo by multiple nano-motors is conceptually analogous to the transport of a green leaf or a dead cockroach by a team of ants \cite{czaczkes13,mccreery14,mccreery16,mccreery17,feinerman18,heckenthaler23,berman11,farivarnejad22} although there are some important differences between the two transport phenomena. Therefore, it is not surprising that the modeling strategies, based on force-sensing, adopted for studying both these problems are quite similar. Theoretical modeling of both these transport phenomena have seen enormous progress in the last 20 years since our review \cite{csk05polrev} summarized the current status at that time; some of the key developments, that are important from the perspective of physics of life, will be mentioned briefly in this article. 

Under varieties of circumstances, many motor proteins move simultaneously along a single filament and this collective movement is referred to as molecular motor traffic because of its apparent similarity with vehicular traffic \cite{nosc05}. Similarly, on a much larger scale, the movement of large number of ants on a pre-existing trail is referred to as ant traffic \cite{burd02}. Both these phenomena have been modelled extensively, although not exclusively, in terms of various adaptations and extensions of Asymmetric Simple Exclusion Process (ASEP), a fundamental model in non-equilibrium statistical physics of self-driven interacting particles \cite{css00,scnbook,gupta23}. Theoretical studies of both these phenomena were in their infancy in 2005; significant progress have been made in the last 20 years. For example, in spite of the fact that the underlying processes are noisy and error prone, biological functions are, to a large extent, fault tolerant. In this article we review, from the perspective of physics of life,  how traffic of molecular machines along the respective filamentous tracks can suppress or enhance errors in several subcellular processes. The interaction between the ants through a pheromone field \cite{theraulaz99,heylighen16a,heylighen16b} created, and sustained, by them has inspired the floor-field model of human pedestrian traffic \cite{burstedde01,kirchner02} which we briefly describe before concluding this article. Thus, we demonstrate how similarities between transport and traffic-like phenomena at different scales have been exploited in the last twenty years for modelling these phenomena thereby enriching the understanding of physics of life.

\section{Intracellular cargo transport along a tubular track: Tug-of-war of opposing motors}

intracellular transport (see \cite{mogre20,sen24} for reviews), which has been an active field of research for last few decades, is also important for several other related phenomena, for example traffic jam in long cell protrusions \cite{pinkoviezky14} and their length control \cite{patra22}. Microtubules, having a tubular structure, serve as tracks for directed stepping of two superfamilies of molecular motors called kinesin and dynein \cite{barlan17}. Similarly, molecular motors belong to the superfamily myosin use actin filaments for cargo transport \cite{titus18}. 
Processive molecular motors that can walk over its track over a long distance without suffering complete detachment from the track are often colloquially referred to as `porters'. Some of the fundamental questions in this context are as follows \cite{mallik13,hancock14}:\\
(i) How many porters are needed to haul a cargo that is at least one or two orders of magnitude larger than an individual motor?\\
(ii) For a given cargo, how is the optimal number of porters determined? \\
(iii) How is the cargo transported in the correct direction, i..e, where it is needed, in spite of the expected tug-of-war between porters whose natural directions of walk on a given filament are oppositely oriented? \\

In the tug-of-war model \cite{muller08} the load is shared among the motors pulling the cargo in the same direction and the detachment of a motor from the cargo is dependent on the load-force directed opposite to its own natural direction of motion on the track. Subsequently, several extensions of this model have been published (see \cite{munoz22} for a more recent extension as well as references for earlier versions). As we will discuss in the next section, force-dependent detachment from the cargo can be a mechanism of collective transport also at a large scale in in the world of living systems. 

Moreover, the cargo need not be a rigid object like a hard sphere; soft cargoes can get deformed by the pull of porters and even result in long tubular protrusions \cite{campas06,leduc10}. An even more fascinating subcellular transport phenomenon is observed in intraflagellar transport (IFT) in eukaryotic cells where the motors collectively pull an IFT train that is formed by the aggregation of multiple IFT particles. These trains carry cargo (usually monomeric constituents of some other proteins). During transport, IFT trains can fuse to form longer IFT trains whereas sometimes a single IFT train can undergo fission resulting in two distinct IFT trains of smaller lengths. ASEP-based model has been developed \cite{patra18} to investigate the effects of fusion and fission of IFT trains on the overall traffic of these trains.

\section{Cargo transport by ants along a temporary trail: emergence of the direction of transport}

Ants \cite{holldoblerAntsbook} are one of the four well known species of social insects \cite{brianSocInsectbook} where social behavior emerged long before man, the social animal, descended on earth. Living as a member of a social insect colony is more subtle than living as a member of a group, for example, of hyenas \cite{krauseGroupbook}. Ants drop a chemical (generically called a pheromone) on the substrate as they crawl forward; although the trail pheromone evaporates the rate of evaporation is slow enough so that other following sniffing ants can pick up its smell and follow the trail while reinforcing it by dropping more pheromones. In this section we do not discuss the phenomenon of self-organization of ant trails \cite{camazineSObook}. The trails are `temporary' in the sense that once the trail pheromones evaporate completely the trail itself disappears. Our interest here is the transport and traffic phenomena that take place on a pre-existing trail. 

Social living in an ant colony does not necessarily imply that these ants transport single large food items to the nest collectively. In case of several species of ants that are otherwise highly social, most of the food is carried to the nest independently by individual ants; if the food item found is too large to be carried by a single ant, it is first dissected into pieces that are small enough to be carried then by single ants. In contrast, some species of ants can collectively transport a cargo that weights more than 5000 times the weight and 10,000 times the volume of a single worker ant without dissecting it \cite{holldoblerAntsbook}. However, the reason for adopting such collective transport may vary from species to species. Aggressive raiding species of ants, like army ants \cite{kronauerArmyAntsbook}, are good examples of collective transporters;  instead of wasting time on in-place dissection of a prey, they simply carry it to the nest collectively, thereby freeing up time and labor for hunting. In contrast, some less aggressive species of ants, like leaf-cutting ants  \cite{holldoblerLeafCutAntsbook,wirthLeafCutAntsbook}
 that are also efficient collective transporter are known to adopt this strategy to rush faster with the food to the nest before any possible competing aggressive species can lay claim on the same food \cite{berman11}. . 

There are several fundamental questions that can be answered only by careful experiments although the observed results may be specific to the type of ant chosen for the experiments. Typical questions are as follows \cite{czaczkes13,mccreery14,mccreery16,mccreery17,feinerman18,heckenthaler23,berman11,farivarnejad22}:\\
(1) So far as an individual ant is concerned, does it push or pull the cargo or simply lift the cargo off the ground and carry it on its back like a porter while walking towards the destination \cite{czaczkes13}? Or, does it internalize the cargo within its body (e.g., in stomach) and spit it out upon reaching destination in the nest? \\
(2) How do multiple ants interact among themselves that result in cooperation and coordination of their actions to transport the cargo in the intended direction (normally, in case of food, towards the nest)? \\
(3) How does the optimum number of ants required for hauling a particular cargo dynamically emerge from the interactions and/or communication among the ants? \\
(4) On the way if a cargo gets stuck in a deadlock how do the ants detect the blockage and then extricate  the cargo from that deadlock?

The major problem of the collective transport of a cargo by multiple ants is how to navigate in the terrain, that is how to move in a desired direction on a relatively short time scale as well as on a longer time scale so as to reach the target site which is, most often, the nest. The ants must collectively carry the cargo in a specific direction; if the cargo is a food item or required for food production in the nest the desired direction of the transport is the nest. However, navigation around an obstacle on the way may require a temporary change of direction of movement for a short duration that would allow bypassing the obstacle \cite{mccreery17}. Such emergent  behaviour \cite{johnsonEmergencebook} displayed in transport phenomena is a classic example of collective animal behavior \cite{sumpterbook} and display evidence of collective intelligence \cite{arimaCollIntelligbook} or, more appropriately in case of ants, swarm intelligence \cite{bonabeauSwarmIntelligbook,garnier07}.

Just as in the case of collective transport of vesicular cargoes by molecular motors, the collective transport of cargoes by ants is also described in terms of dynamical  equations in the overdamped limit where the velocity (rather than acceleration) is proportional to the externally applied force. Therefore, an average velocity of the cargo in the direction of the nest is possible only if the resultant force is also in the same direction. If the movement of the individual porter ants are mutually {\it uncoupled} then to accomplish the task of transportation of the cargo the ants need to be aware of the direction of the nest and exert force in that direction. A possible model for achieving this is the classic ``many-wrongs principle'' underlying group navigation in many species of migrating birds \cite{simons04,codling07}. An alternative mechanism assumes that the individual ants do not knowingly exert force in the desired direction (the nest, for example); instead, they `sense' the current direction of motion of the cargo and align the force they individually exert in that direction. The movement of the cargo in the `correct' direction emerges from the cargo-mediated mechanical coupling of the porter ants (see \cite{feinerman18} for details and review). 

One of the earliest mathematical models of collective transport by ants was reported by McCreery et al. \cite{mccreery16,mccreery17}. In this model, at any instant of time, each ant is assumed by in one of the three allowed states: (1) disengaged from the cargo (D), (2) trying to move the cargo to the left (L), 2) trying to move the cargo to the right (R); the number of ants in the states D, L and R at time $t$ are denoted by $N_{D}, N_{L}$ and $N_{R}$, respectively. Direct transition from D to L (with rate $W_{D \to L}$) and that from D and R (with rate $W_{D \to R}$) and the corresponding reverse transitions (with the corresponding rates $W_{L \to D}$ and $W_{R \to D}$) are allowed, no direct transition between L and R is allowed. Different functional forms of dependences of the rates of these transitions on the number of ants in the states $L$ and $R$ were assumed in different plausible behavioral scenarios. Ordinary differential equations are written to describe the deterministic time evolution of the population in these three states; a stochastic extension of this dynamics was also reported (see \cite{mccreery16,mccreery17} for the details). 

Any directional bias where $L \neq R$ would result in a directional movement of the cargo; the symmetry in the two directions on the trail is hereby broken trivially by the directional bias in the kinetics. In contrast, in the case of unbiased rates, i.e., $L=R$, a deadlock results, as expected, irrespective of the behavioral model assumed for the ants. The deadlock remains stable if the dynamics is deterministic. Although, on the average, the dynamics of the cargo in the stochastic model are mostly similar to the corresponding deterministic dynamics, there is one key difference: the noise in the stochastic model can break the deadlock even in the symmetric case $L=R$ and select directional movement provided the ants exhibit ``informed'' behavior. The `informed' behavioral rule was defined as follows: {\it If an ant is in one of the two possible engaged states, and if it senses that the transport is successful, then it disengages (a) readily if it finds itself opposing the majority, and (b) less readily if it finds itself going along the same direction as that of the majority}. 

Thus, sensing the direction of movement of the majority of the ants, or that of cargo, is a key feature of the coupled porters model. At present, mechanical coupling through cargo-mediated force-sensing seems to the most popular and plausible scenario \cite{feinerman18} that has similarities with the models of collective transport of vesicular cargoes by molecular motors. However, one key difference is that unlike an ant, which can pull or push the cargo in any direction, kinesin and dynein motors can walk only in the plus and minus ends of the microtubule track and not in the opposite direction.

\section{Effects of traffic on a tortuous template: regulation of search and errors in gene expression} 

Genetic message is encoded chemically in the sequence of nucleotides, the monomeric constituents of a linear heteropolymer called DNA. The four letters of the alphabet of this chemical language are represented by the symbols A, T, C, G. This genetic message is transcribed into another message, encoded also chemically in the sequence of the monomeric subunits of a linear heteropolymer called RNA; the language of encoding on RNA use (slightly different)  four letters A, U, C, G. The process of transcription is carried out by a molecular machine \cite{frank10} called RNA polymerase (RNAP) \cite{bucRNAPbook}. During error-free transcription, a RNAP would move unidirectionally along the DNA template in a step by step manner; in each step it moves forward by one nucleotide on the template DNA and elongates the growing RNA by one nucleotide that is complementary to the corresponding nucleotide on the template DNA  (A and U are complementary to each other just as C and G are). When large number of RNAs are to be synthesized within a short duration, several RNAPs begin transcribing the same DNA template in quick succession; their collective movement on the DNA strand is often referred to as RNAP traffic \cite{tripathi08} because of the superficial similarity with vehicular traffic on highways \cite{css00,scnbook}.

Amino acids are the monomeric subunits of proteins each of which is a linear polymer.  In any given protein, the particular sequence of the types of amino acids is determined by the sequence of nucleotides, the monomeric subunits, of the corresponding template messenger RNA (mRNA) molecule. The polymerization of a protein from scratch, as directed by the corresponding mRNA template, is carried out by a molecular machine, called ribosome \cite{frank10,Chetverin21}. Normally, the amino acids can be of 20 different types whereas the nucleotides on a mRNA can be of only 4 different types. Therefore, the template-directed polymerization of a protein is called translation whereby the genetic message encoded on the mRNA using a 4-letter alphabet gets translated to another that uses a 20-letter alphabet. Often large number of ribosomes translate the same mRNA simultaneously; their spatio-temporal organization resembles vehicular traffic on the mRNA template which serves as the track for their motor-like movement. Historically speaking, TASEP was introduced originally as an attempt to model ribosome traffic on mRNA (see ref.\cite{haar12,shyam24} for history and review of the older works); it has been extended in more recent times incorporating newly discovered aspects of translation 
\cite{basu07,garai09,sharma11}.

\subsection{Effects of RNAP traffic on target site search by diffusing transcription factors: Brownian ratchet?} 

Several important biological processes are initiated by the binding of a protein to a specific site on the DNA; the transcription factors (TF)  are among the most prominent examples of such proteins. The search strategies and the effects of static roadblocks, like DNA-bound proteins, on the average time needed for finding the target have been investigated over several decades (see ref.\cite{park21,iwahara21} for reviews). If RNAP traffic flows on a segment of DNA transcribing it while, simultaneously, a TF diffusively searches for a 
target site on the same segment, the two processes can affect each other. Motivated by this phenomenon, Ghosh et al.\cite{ghosh18b}  developed a kinetic model where a ‘particle’, that represents a TF, diffuses on a one-dimensional lattice searching for a specific site on it. On the same lattice another species of particles, each of which represents a RNAP, hop from left to right exactly as in a TASEP. Simultaneous occupation of any site by more than one particle, irrespective of their identities, is forbidden. An empty site, irrespective of its location on the lattice, can can get occupied by a fresh TF. Similarly a TF that already occupies a site can get detached irrespective of the location of the site on the lattice. In contrast, the RNAPs can attach only to the first site at the left edge and detach from only the last site on the right edge of the lattice. By a combination of approximate analytical calculations and computer simulations, Ghosh et al.\cite{ghosh18b} showed that RNAP traffic rectifies the diffusive motion of a TF to that of a Brownian ratchet \cite{reimann02,cuberoRatchetbook,hoffmannRatchetbook,lau17} and, consequently, the mean time of successful search by the TF can be even shorter than that needed in the absence of RNAP traffic.

\subsection{Traffic of RNAPs transcribing two different genes simultaneously: Transcriptional Interference} 

\begin{figure}[t] 
\begin{center}
\includegraphics[width=0.5\columnwidth]{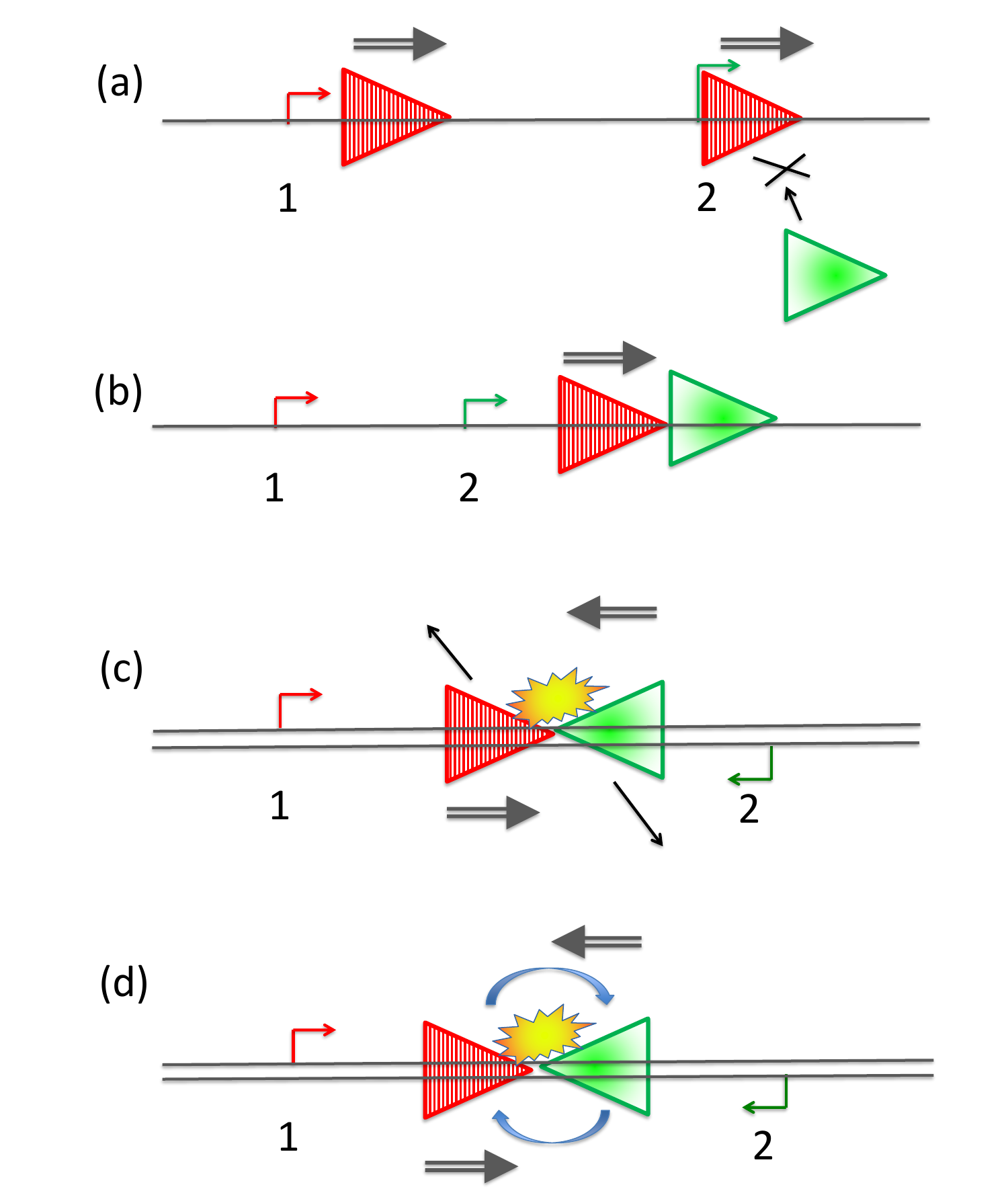} 
\end{center} 
\caption[scale=tiny]{(Color online) Schematic representation of the common modes of transcriptional interference (TI) \cite{shearwin05,mazo07}: (a) occlusion (in co-directional TI), (b) road block (in co-directional TI), (c) head-on collision (in contra-directional TI) resulting is detachment of the RNAPs, (d) head-on collision (in contra-directional TI) resulting in the RNAPs passing each other. Single step-like arrows labelled by 1 and 2 indicate the sites of initiation of transcription of the genes 1 and 2. The two distinct species of RNAP are represented by shaded red and open green triangles where the horizontal orientation of the triangle indicates its natural direction of movement. Each of the double arrows indicates the direction in which the corresponding RNAP has a natural tendency to move at that instant of time; absence of double arrow implies that the RNAP is stalled at that moment. The single straight arrows in (c) indicate the possible detachment that results from head-on collision whereas the semi-circular arrows in (d) depict the passing of the RNAPs while approaching each other head-on. (Fig.1 of ref.\cite{ghosh16})} 
\label{fig-TI}
\end{figure}

\begin{figure}[t]
(a)\\
  \includegraphics[angle=0,width=0.5\columnwidth]{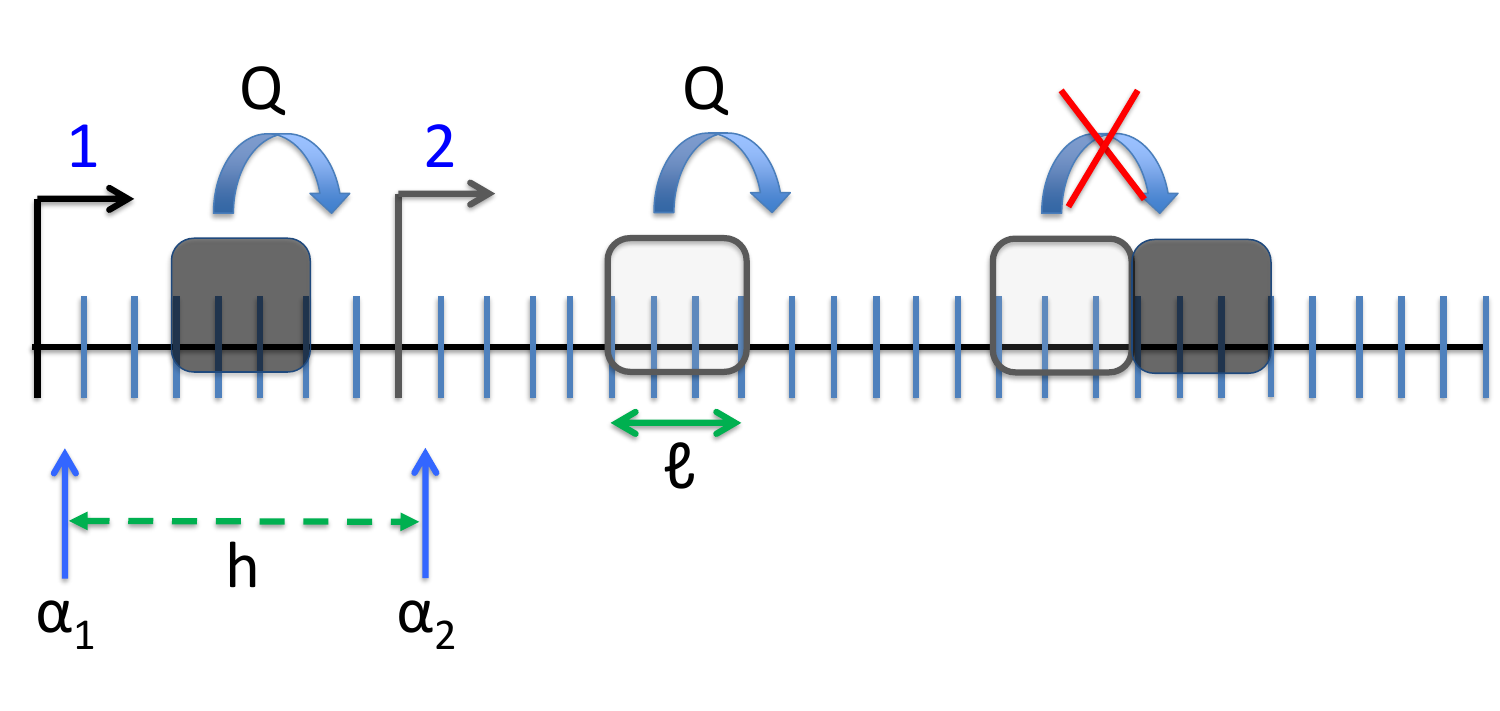} \\
(b) \\
  \includegraphics[angle=0,width=0.5\columnwidth]{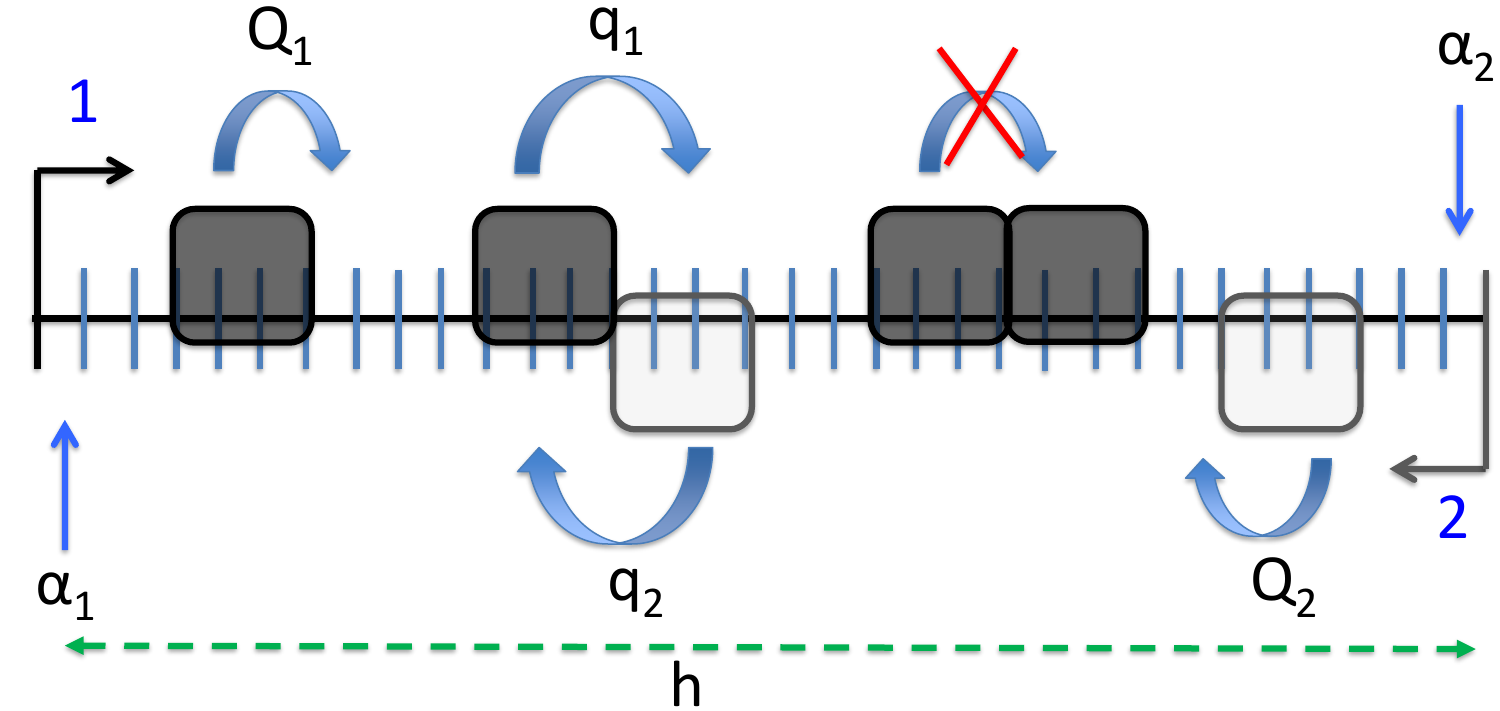} \\
(c) \\
  \includegraphics[angle=0,width=0.5\columnwidth]{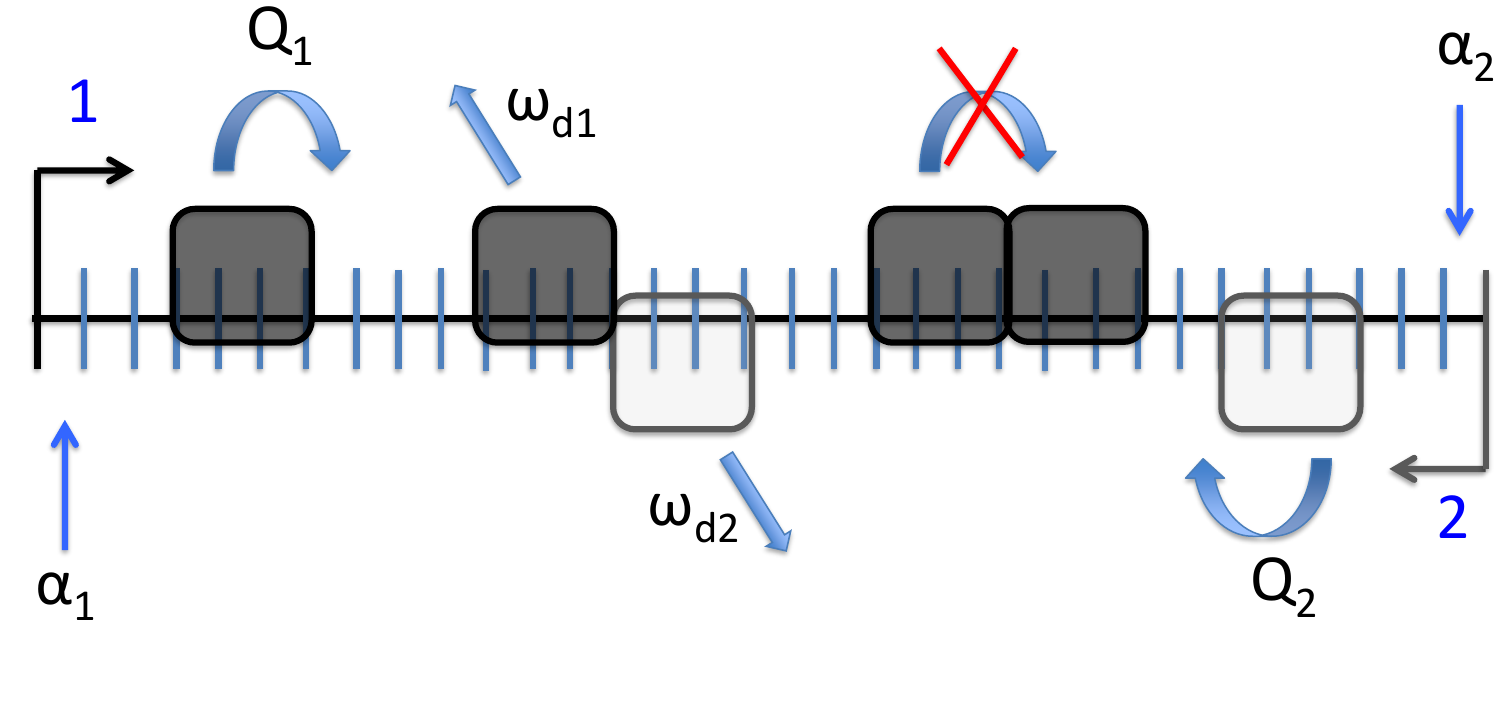} \\
  \caption{(Color online) Schematic representation of the kinetics of the two-species exclusion model \cite{ghosh16} that was motivated by Transcriptional Interference (TI) sketched in Fig.\ref{fig-TI}. A single lattice is used for formulating the models for both co-directional (in(a)) and contra-directional (in(b)-(c)) movement of the 
two species of rigid rods represented by filled and open rectangles.  The ON-ramps are marked by the step-like arrows labelled by 1 and 2; the OFF-ramps are also distinct (but not shown explicitly), in general. In all the cases (a)-(c) the members of the same species of rods use the same ON-ramp and carry the label of the ON-ramp (1 or 2) throughout their journey. $\alpha_{1}$ and $\alpha_{2}$ are the rates of initiation (i.e., entry) of the two species of rods at their respective ON-ramps. The different hopping rates in the bulk under different conditions are  also shown along with the corresponding arrows. The symbol ${\ell}$ denotes the linear  size of each hard rod of both the species while the separation between the two ON-ramps is denoted by $h$ (Fig.2 of ref.\cite{ghosh16}). 
}
  \label{fig-TImodel}
\end{figure}


Ghosh et al. \cite{ghosh16} developed a multi-species exclusion model inspired by a biological phenomenon called transcriptional interference (TI) \cite{shearwin05,mazo07}. This interference arises from simultaneous transcription of two overlapping genes \cite{wright22} encoded on the same DNA template or two genes encoded on the two adjacent single strands of a duplex (double-stranded) DNA (see Fig.\ref{fig-TI}). In the former case RNAP traffic is entirely uni-directional, as sketched in Fig.\ref{fig-TI}(a) and (b), although RNAPs transcribing different genes polymerize two distinct species of RNA molecules by starting (and stopping) at different sites on the same template DNA strand. In the case of co-directional TI both the genes may share a common termination site (off-ramp); that situation is captured as a special case of the more general formulation by Ghosh et al.\cite{ghosh16}. In the case of contra-directional TI, as sketched in Fig.\ref{fig-TI}(c) and (d), RNAP traffic in the two adjacent ‘lanes’ move in opposite directions transcribing the respective distinct genes. If the sites of termination of both the transcriptional processes are outside the region of the overlap of the two genes encoded on the two adjacent DNA strands the arrangement is defined as ‘head-to-head’. In contrast, in the ‘tail-to-tail’ arrangement of the genes the sites of initiation marked on the two adjacent strands of the duplex DNA are beyond the region of overlap of the two genes. In both these situations the two interfering transcriptional processes have suppressive effects on each other \cite{shearwin05,mazo07,wright22}.

Ghosh et al.\cite{ghosh16} introduced exclusion models (see Fig.\ref{fig-TImodel}) of two distinguishable species of hard rods with their distinct sites of entry and exit on the lattice under open boundary conditions. In the first model both species of rods move in the same direction whereas in the other two models they move in the opposite direction. The specifications of the entry and exit sites as well as the rules for the kinetics of the models, particularly the rules for the outcome of the encounter of the rods, were formulated to mimic those observed in TI \cite{shearwin05,mazo07}. 
By a combination of mean-field theory and computer simulation of these models Ghosh et al.\cite{ghosh16}  demonstrated that it is possible to switch off the transcription of one of the two genes by the ongoing transcription of the other. Moreover, by a detailed exploration of the three scenarios encapsulated by Fig.\ref{fig-TImodel} they also discovered more than one possible mechanisms of the switch-like regulation of the fluxes.

\subsection{Conflict between RNAP Traffic and Replication fork: Simultaneous Transcription and DNA replication} 

\begin{figure}[h]
    \begin{center}
        \includegraphics[angle=0,width=0.5\columnwidth]{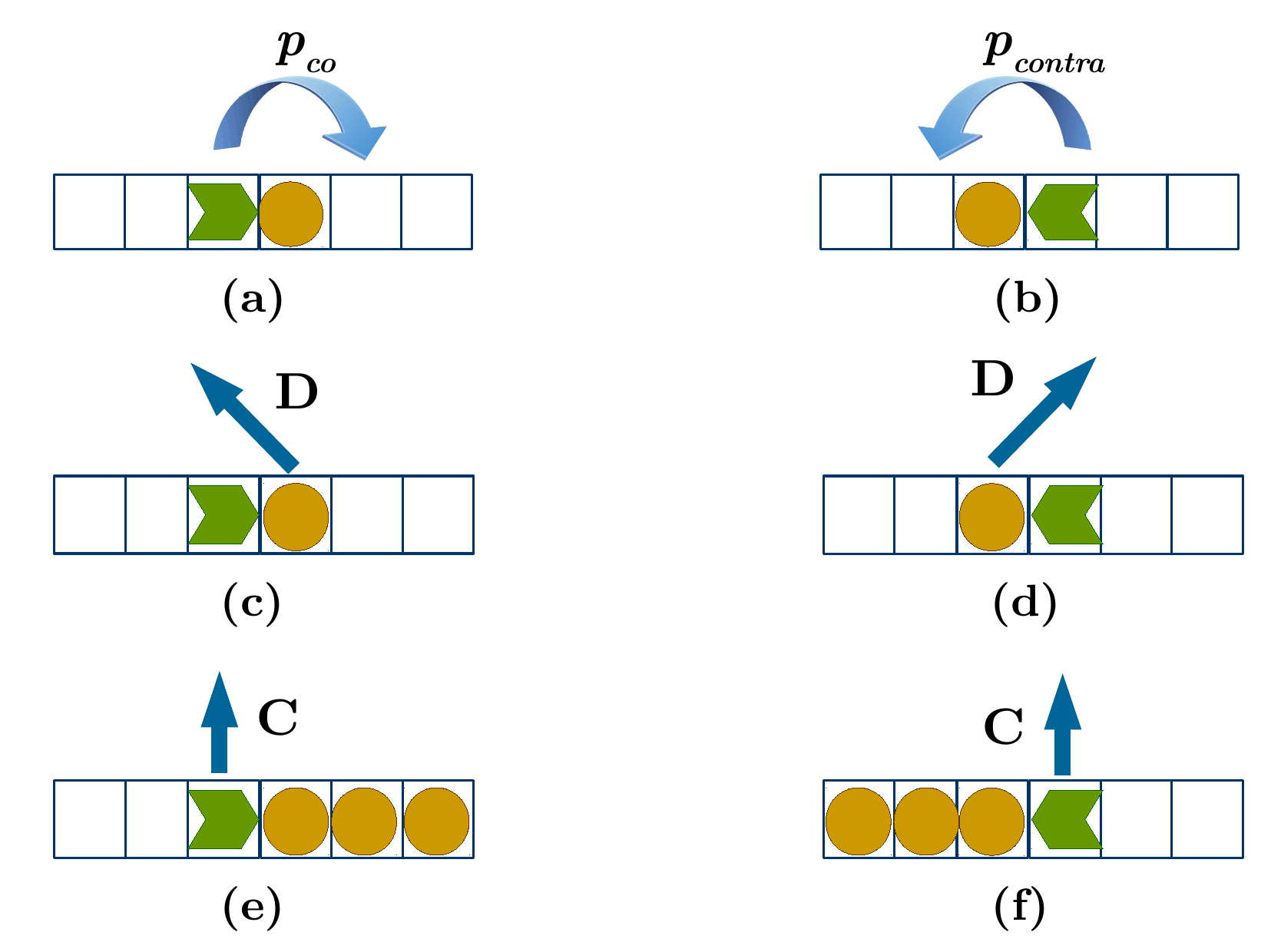}\\[0.02 cm]
    \end{center}
    \caption{Schematic depiction of resolution of conflict between the replication (fork), represented by $R$ particles, and transcription, represented by $P$ particles. In (a) and (b) $R$ particle can pass the $P$ particle, with rates $ p_{\text{co}} $ and $ p_{\text{contra}}$, respectively; both replication and transcription continue after a brief slowdown during passing. In (c) and (d) the $R$ particle can knock the $P$ particle out of the track, with rate $ D $ in both the situations irrespective of the direction of encounter; replication continues while transcription is aborted prematurely. In (e) and (f) $P$ particle can block the progress of the $R$ particle thereby eventually causing catastrophic collapse of replication, with rate $ C$. (Fig.2 of ref. \cite{ghosh18})}
    \label{fig-model-bypass}
\end{figure}

Ghosh et al. \cite{ghosh18} introduced a two-species exclusion model to describe the key features of the conflict between the RNA polymerase (RNAP) motor traffic, engaged in the transcription of a segment of DNA, concomitant with the progress of two DNA replication forks on the same DNA segment \cite{pomerantz10}. One of the species of particles (P) represents RNAP motors while the other (R) represents the replication forks. Motivated by the biological phenomena that this model is intended to capture, a maximum of two R particles only are allowed to enter the lattice from two opposite ends whereas the unrestricted number of P particles constitutes a totally asymmetric simple exclusion process (TASEP) in a segment in the middle of the lattice. Mutual exclusion is captured by the simple rule that no site can be occupied simultaneously by more than one particle irrespective of the species to which it belongs. 

Three possible pathways for the resolution of conflict between transcription and replication (shown schematically in Fig.\ref{fig-model-bypass}) are as follows: \\
(a) The R and P particles can bypass each other, irrespective of co-directional or contra-directional encounter, without dislodging either of the two from the lattice;\\
(b) The R particle can knock the P particle out of the track, thereby aborting the transcription by that particle P particle prematurely while the replication resumes and continues;\\
(c) The P particle detaches from the lattice resulting in collapse of the replication fork while the P particle resumes transcription. 

By a combination of analytical arguments and computer simulation, Ghosh et al. \cite{ghosh18} demonstrated how the RNAP traffic affects not only the DNA replication time but also the frequency of unsuccessful replication events because of transcription-replication conflict.

\subsection{Backtracking of RNAP and correction of random transcriptional errors: effects of RNAP traffic} 

It has been known for a long time \cite{kirkwoodErrorbook} that not only are the chemical processes of RNA elongation as well as the mechanical process of forward stepping of the RNAP intrinsically stochastic the selection of the monomeric subunits for the synthesis of the RNA are also error prone. Out of the $10^5$ nucleotides selected sequentially, typically, one would be wrong. Interestingly, nature also has mechanisms of correcting such `random errors'. Detecting the incorporation of a wrong monomer at the growing tip of the RNA, the RNAP `backtracks' by a few nucleotides through diffusion on its template DNA and pauses there. Subsequently, two pathways are open for it: (i) it can either diffusively return to position from where it backtracked and resume transcription without correcting the transcriptional error, or (ii) during its pause at the backtracked position a specialized enzyme engages with the extruding tip of the RNA and catalyzes the cleavage of misincorpared (non-complementary) nucleotide after which RNAP can resume transcription beginning with the incorporation of the correct nucleotide in place of the earlier error. When the latter pathway is followed, the mode of error correction is usually referred to as transcriptional `proofreading' (see ref.\cite{nudler12} for a review on backtracking, its causes and consequences).  

What happens if a RNAP that is on the verge of backtracking finds inadequate vacant space upstream on the template DNA? {\it In-vitro} and {\it in-vivo} experiments \cite{epshtein03a,epshtein03b} have  convincingly demonstrated that backtracking of a RNAP gets strongly suppressed if it is closely followed by a trailing RNAP because there is practically no vacant space available on the DNA track for the backtracking. Can such a situation arise really under physiological conditions of a cell? If the traffic of RNAPs is not highly congested individual RNAPs may not encounter any difficulty in backtracking thereby allowing for the possibility of transcriptional proofreading. But, in a dense traffic backtracking, and consequently proofreading, can be suppressed significantly resulting in a relatively low fidelity of transcription \cite{klumpp11}. This possibility has been supported by the results of analysis of TASEP-based models of RNAP traffic that allow for backtracking and the pathways for resumption of transcription by backtracked RNAPs \cite{sahoo11,sahoo13}.

\subsection{Slippage of transcript in a RNAP as a `programmed' transcriptional error: effects of RNAP traffic} 

Experimental investigators have discovered the existence of specific stretches of some DNA sequence where, during transcription, the RNAP may lose its `grip' on the growing RNA which, consequently slips backward with respect to the RNAP. However, during the slippage of the transcript, the RNAP motor itself does not slip simultaneously on its DNA track \cite{anikininbook,atkins16}. At any such site where the transcript is slippage-prone multiple successive events of backward slippage may occur before the transcription process resumes. Thus, backward transcript slippage results in longer final products of transcription because of the incorporation of more nucleotides as compared to the length of the transcript that would have been synthesized in the absence of transcript slippage. This error of transcription does not occur at any arbitrary site randomly, but `programmed' in the sequence of the monomeric subunits of the DNA template. The detailed mechanism of this programmed error is still under intense investigation. 

In a congested traffic each RNAP is expected to dwell, on the average, longer at a site, including those sites where the nascent RNA remains slippage prone. The longer a RNAP dwells at the site where the nascent RNA is slippage prone, the larger would be the number of transcript slippage events it is likely to suffer. Thus, traffic congestion can result in longer transcripts than those resulting from transcript slippage by an isolated RNAP motoring unhindered along the same DNA template. This effect of RNAP traffic on transcript slippage frequency has been demonstrated explicitly by a TASEP-based model that was developed by Ghosh et al \cite{ghosh19} to investigate the interplay of transcript slippage and RNAP traffic. Transcript slippage, in turn, affects the RNAP traffic because the transcript slippage site on the DNA template acts effectively like a `defect' site in the TASEP model. Consequently, the steady-state spatial distribution of the RNAPs display the known characteristic features of the density profile of particles in TASEP in the presence of defects on the lattice \cite{ghosh19}.

\subsection{Shift of Reading Frame of a Ribosome as a `programmed' translational error: effects of Ribosome traffic} 

\begin{figure}[t] 
\begin{center}
\includegraphics[width=0.5\columnwidth]{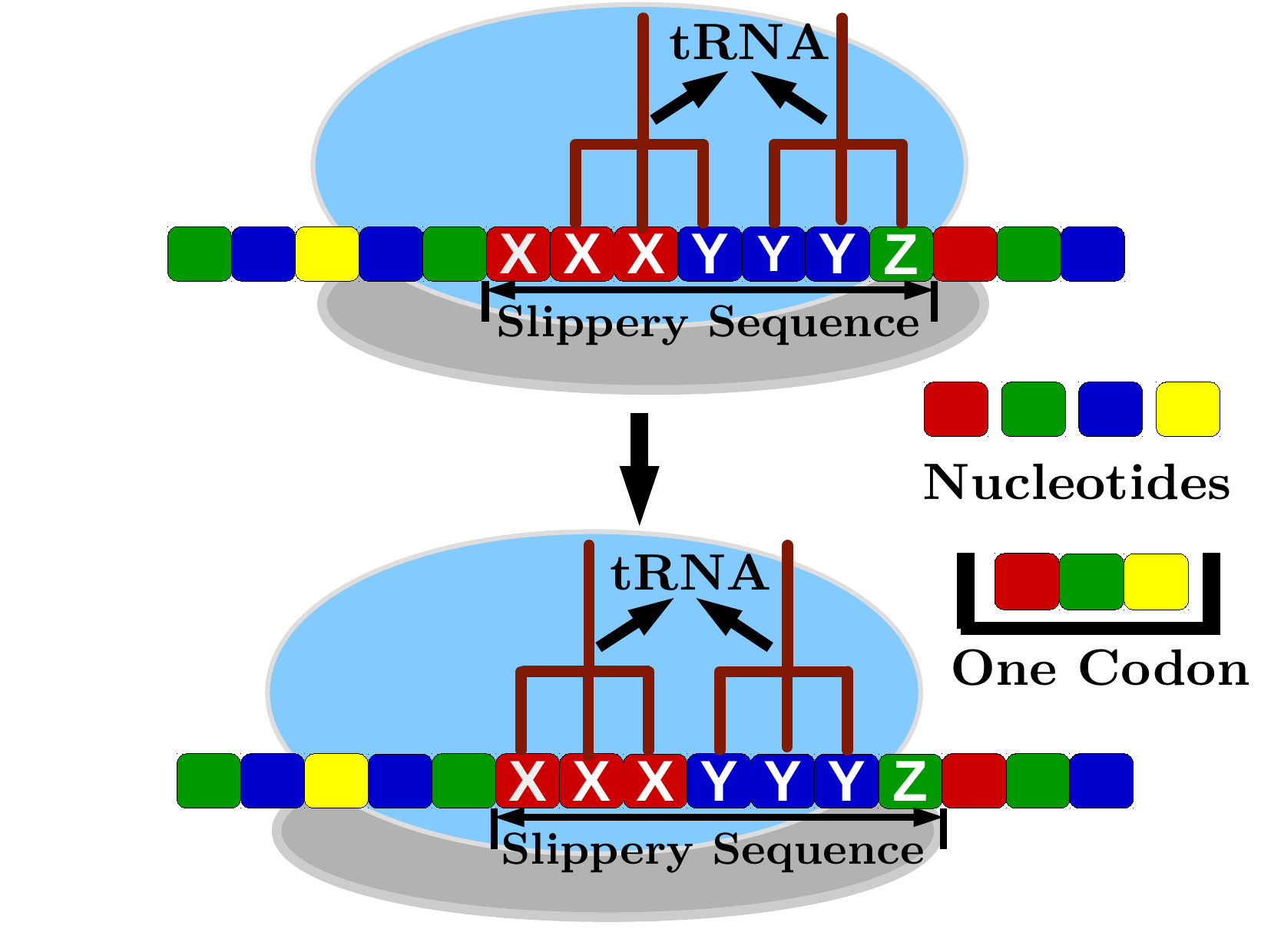} 
\end{center} 
\caption[scale=tiny]{(Color online) A schematic depiction of -1 programmed ribosomal frameshift (PRF) from initial reading frame to -1 reading frame. The chain consisting of rectangles of four colors (red, green, blue, yellow) represent a sequence of four types of nucleotides, the monomeric subunits of the mRNA template. The bulky object consisting of the light blue and grey parts represents a ribosome (Fig.1 of ref. \cite{msc16}). } 
\label{fig-kinetics}
\end{figure}

Strictly speaking, translation involves myriads of devices and accessories as well as coordination of a large number of well orchestrated sub-processes. However, the three major stages into which the overall process is divided are (a) initiation, (b) elongation, and (c) termination. Elongation of the growing protein by a ribosome takes place in a step-by-step manner: the incorporation of each amino acid monomer at the growing tip of the protein is accompanied by a forward stepping of the ribosome on its mRNA template by one codon, each codon being a triplet of nucleotides. Which codon corresponds to which amino acid is dictated by the genetic code. 
Interestingly, for a ribosome engaged in polymerizing a protein, the mRNA template also serves as a track for its its forward movement by three nucleotides in each step; the directed movement of the ribosome is driven by the simultaneous conversion of chemical energy into mechanical work. Thus, in analogy with cytoskeletal motor proteins, a ribosome is also regarded as a molecular motor \cite{chowdhury13,kolomeiskybook}. 

From our perspective adopted here, ribosome is a motor that steps forward probabilistically by one codon in each step on a tortuous linear track. At each codon the ``reading frame'' of the ribosome decodes the message encoded in that triplet of nucleotides of the mRNA template and then slides to the next codon. This reading frame is established in the initiation stage of translation. This reading frame is maintained faithfully as the ribosome hops from one codon to the next during the course of normal elongation of the protein. In other words, normally the ribosome neither skips any nucleotide on the mRNA nor reads a nucleotide more than once. However, on many template mRNA strands there are some special ``slippery'' sequences of nucleotides where a ribosome can lose its grip on its track, resulting in a shift of its reading frame either backward or forward by one or more nucleotides. These processes are referred to as ribosomal frameshift \cite{atkins16,brierleybook}. 

The most commonly occurring, and extensively studied, cases correspond to a shift of the reading frame backward or forward by a single nucleotide on the mRNA track. These are referred to as -1 frameshift and +1 frameshift, respectively. After suffering a programmed frameshift, the ribosome resumes its operation but, from then onwards,  decodes the mRNA template using the shifted reading frame (see Fig.~(\ref{fig-kinetics})). The sequence of codons read by the shifted reading frame is, in general, different from the sequence that would be read if the ribosome had not suffered frameshift. Thus, a frameshift produces a `fusion' protein; a classic example of such a fusion product of -1 frame shift is the gag-pol fusion protein of the human immunodeficiency virus (HIV) \cite{atkins16,brierleybook}. Programmed frameshift, is a well known mode of non-conventional translation \cite{firth12} and it happens to be one of the several modes of more general phenomenon of genetic recoding \cite{atkins10}.

For a long time it was established fact that programmed -1 frameshift requires two key ingredients: (a) A `slippery' sequence (usually about seven nucleotide long) on the mRNA, and (b) a secondary structure (usually a pseudoknot \cite{atkins16,brierleybook}) of the mRNA  that is located about 6 nucleotides downstream from  the slippery sequence. The pseudoknot can act as a roadblock  against the forward movement of the ribosome which is, therefore, forced to make a long pause on the slippery site where the ribosome is most likely to suffer slippage. One question that did not attract wide attention of the investigators till 2016 is: what is the role of the `stiffness' of the pseudoknot on the frequency (or probability) of frameshift at the slipper sequence immediately upstream from the pseudoknot? This question was addressed, to our knowledge, for the first time by Mishra et al. \cite{msc16}; the answer is related to the well known phenomenon of traffic-like collective movement of ribosomes on a mRNA template \cite{csk05polrev,chowdhury13} during the elongation stage of translation \cite{schutz20}. 

In contrast to the earlier TASEP-based models of translation \cite{csk05polrev},  the model introduced by Mishra et al. \cite{msc16}  treated individual nucleotides, rather than triplets of nucleotides (codons), as the basic unit of the mRNA track. This modification was necessary to model -1 frameshift whereby the ribosome slips backward by one nucleotide on its mRNA track. The qualitative trends of the quantitative results obtained by Mishra et al. \cite{msc16} is consistent with what is expected based on physical intuition. If the pseudoknot is very soft then a ribosome can unwind it easily without trying too long to suffer a frameshift. At the opposite extreme is the situation where the pseudoknot is so stiff that passage through this effective barrier becomes extremely slow causing `bumper-to-bumper' congestion. In such a situation, even if a ribosome on the slippery sequence is on the verge of a frameshift, its actual shift is prevented because of the occupation of the codon behind it by the trailing ribosome. Such suppression of -1 programmed frameshift by a trailing ribosome in a dense ribosome traffic on a mRNA track \cite{msc16}, is similar to the suppression of diffusive backtracking of a RNA polymerase (RNAP) motor by another RNAP trailing very closely on a DNA track \cite{sahoo11} that has been reviewed above.

\subsection{'Leaky' scanning by ribosomal subunit and overlapping genes: co-directional encounter and interference} 

Ribosomal frameshift, as explained above, occurs during the stage of elongation of the protein. However, this is not the only known non-canonical translation \cite{firth12}; there are other non-canonical translational processes that involve translation initiation. Loosely speaking, let us call the segment of mRNA in between a start codon and the corresponding stop codon a “gene.” If two genes (say, $G_1$ and $G_2$) that are encoded on the same mRNA overlap, then the rates of synthesis of the corresponding proteins ($P_1$ and $P_2$) would be influenced by the interference of the traffic of the two species of ribosomes engaged in their synthesis. This translational interference would be somewhat similar, at least superficially, with transcriptional interference that we have already discussed above. What complicates the matter even further is that pre-assembled functionally competent ribosomes do not bind directly to the site of initiation of protein synthesis. Instead, one of the major components of a ribosome called the small subunit (SSU), first binds a site upstream from the start site on the mRNA. Then, it searches for the specific start site on that mRNA by scanning along the latter’s contour. Once the SSU reaches the start site and binds it specifically, then another major component of the ribosome, called the large subunit (LSU), binds with the SSU, along with some other accessory molecules, thereby completing the assembly of the ribosome after which normal synthesis of the protein can begin. 

However, a fraction of the SSUs scanning for the first initiation site may miss it because of “leaky” scanning \cite{firth12} and continue scanning beyond that site while the remaining fraction not only binds with the first initiation site but also proceed with the subsequent steps resulting in normal step-by-step elongation of the corresponding protein $P_1$. In contrast, those SSU that miss the first initiation site continue moving co-directionally with the ribosomes translating $G_1$ thereby interfering with the synthesis of the protein $P_1$ until they reach the second start site. A fraction of these SSUs may correctly identify the second start site and assemble into a ribosome thereby becoming competent to begin synthesizing a protein $P_2$. The ribosomes engaged in the synthesis of $P_1$ and $P_2$ continue suffering co-directional encounter and interfere in their function till one of these reach the corresponding site of termination of translation. The SSUs that fail to identify both the first and second sites of initiation continues futile scanning and thereby interfering with the translation of $G_1$ and $G_2$. An exclusion model with three interconvertible species of hard rods was developed for studying the effects of leaky scanning and overlapping genes on protein synthesis \cite{mishra19}.

\section{Ant traffic on existing trails: effects of evaporating pheromone and lane-changing} 

\subsection{Unidirectional single-lane traffic of ants {\it  Leptogenys processionalis} on a natural trail: high-flux puzzle} 

\begin{figure}[t] 
\begin{center}
\includegraphics[width=0.5\columnwidth]{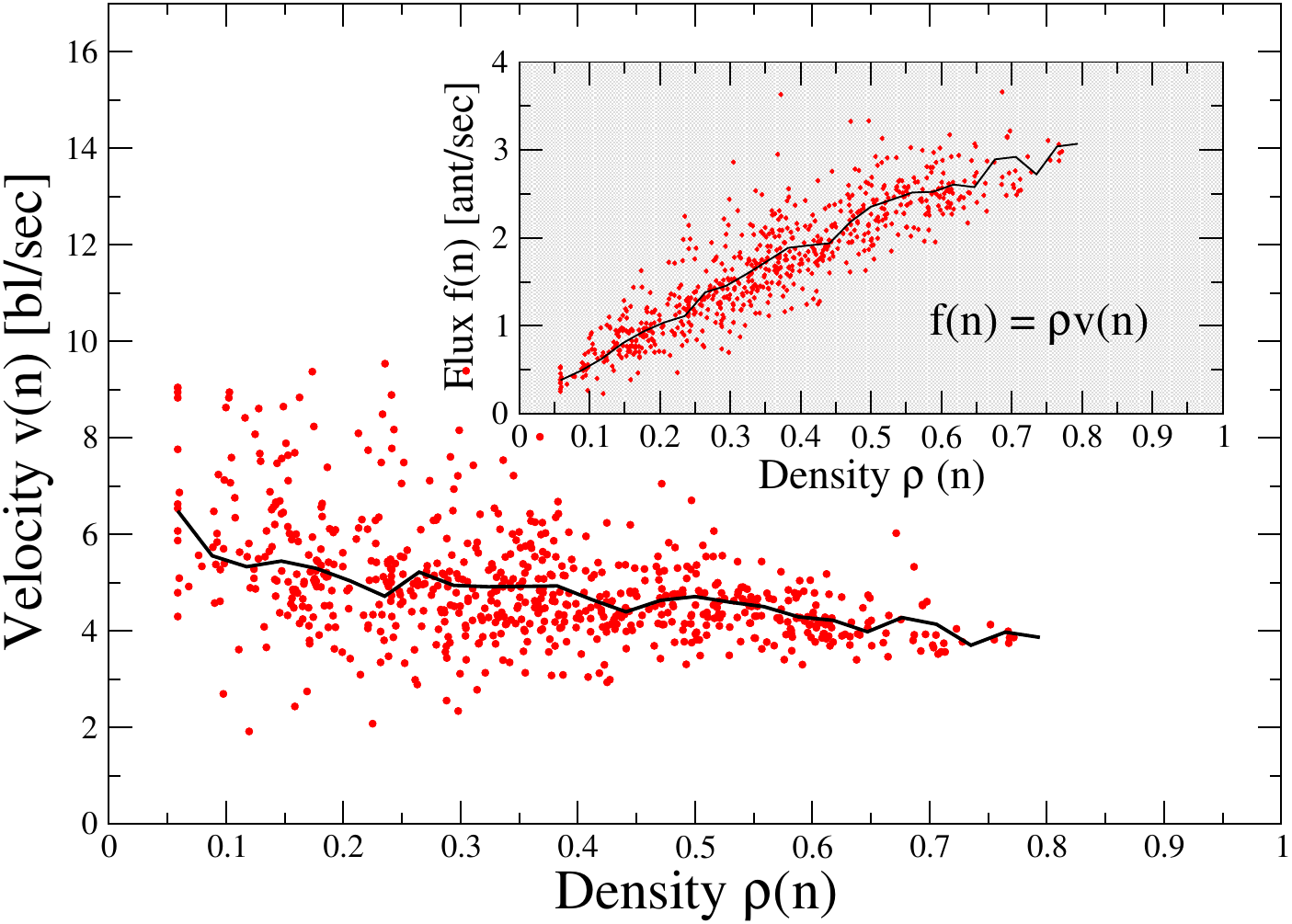} 
\end{center} 
\caption[scale=tiny]{(Color online) Individual velocities (dots) of the ants in a unidirectional traffic along a single-lane trail section of length $L=17~$bl are plotted against the density. The corresponding average velocity, shown by the solid line, is almost independent of the density thereby indicating that mutual hindrance is practically absent even at densities as high as $70\%$.  Consequently, the flux increases almost linearly with the density in the so-called {\it fundamental diagram} plotted in the inset (Fig.3 of ref.\cite{john09})} 
\label{fig-johnexpt}
\end{figure}

Inferring the traffic rules from empirical observations and controlled experiments have always been fascinating \cite{john09,Fourcassie10}. 
One of the surprising empirical results \cite{john09} of single-lane uni-directional traffic of ant species {\it  Leptogenys processionalis} on natural trails is that no jamming occurs at number densities as high as $70\%$. This is sharp contrast to the TASEP-based models of ant-traffic \cite{chowdhury02,nishinari03}. 
Several attempts have been made in the last fifteen years to explain the physical origin of the unusual qualitative features of the results reported in ref.\cite{john09}. One of these is a phenomenological theory \cite{chaudhuri15} where the dynamics of each ant is formulated in terms of Langevin equations in continuous space and time. The mutual exclusion is implemented through a strongly repulsive nearest-neighbour interaction. The self-propelled nature of the ants is captured through a self-propulsion force. The key assumption of this model is that the self-propulsion force acting on an ant adapts its strength with the changing distance headway in front of the ant. 

Whether an ant senses the distance to the closest ant in front directly through limited vision, or touch remains a matter of speculation. Besides, since the model does not capture the effects of pheromone, the implication seems to be that pheromone-mediated communication does not play any important role in the traffic of this ant species. However, subsequently, a pheromone-controlled model has been proposed \cite{guo18} in which the ants adapt their `desired' velocities depending on the concentration of pheromone sensed in the immediate front position on the trail. This seems to be a more realistic account of the adaptive chemotactic motility of the ants. However, more recently alternative theories have been published that claim to reproduce the qualitative features of the results reported in ref.\cite{john09} by appropriately modifying the TASEP-based minimal model developed originally by Chowdhury et al.\cite{chowdhury02} capturing the effects of pheromone. 

More experiments with other species of ants is necessary to understand what traffic rules are adopted by different species of ants to maximize the flow. For example, 
individual leaf-cutting ants {\it Acromyrmex crassipinus} move slower while carrying larger leaves as cargo alone. It has been observed \cite{pereyra20} that these leaf-cutting ants regulate the size of the cargo carried by individual ants for maximizing the flow: large-size cargo leaves are transported mainly under low flow conditions whereas the proportion of such large cargoes were found to decrease with increase in the flow. This is similar to the restrictions imposed on heavy vehicles during peak hours of vehicular traffic. Strong confinement by the walls of a tunnel can give rise to novel phenomena in ant traffic \cite{gravish15}; these 
are likely to be very important in subterranean ant traffic where the ants need to adapt to the confining environment 
so as to avoid clogging the tunnel.

\subsection{Bidirectional multi-lane traffic of ants {\it Eciton burchelli} on existing trails: effects of lane-changing} 

\begin{figure}[t] 
\begin{center}
\includegraphics[width=0.5\columnwidth]{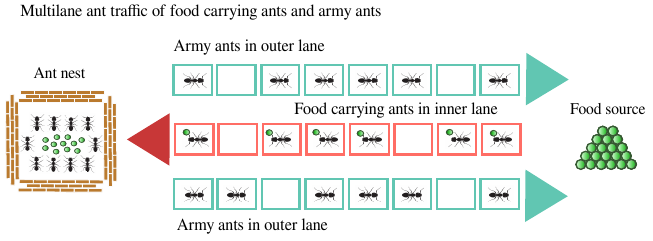} 
\end{center} 
\caption[scale=tiny]{(Color online) Army ants travel left to right in the two outer lanes from their nest
to the food source. Their nest mates that carry food from the source of food to the nest
move right to left along the inner lane. (Fig.3 of \cite{pradhan21})} 
\label{fig-Bidor3LaneModel1}
\end{figure}

\begin{figure}[t] 
\begin{center}
\includegraphics[width=0.9\columnwidth]{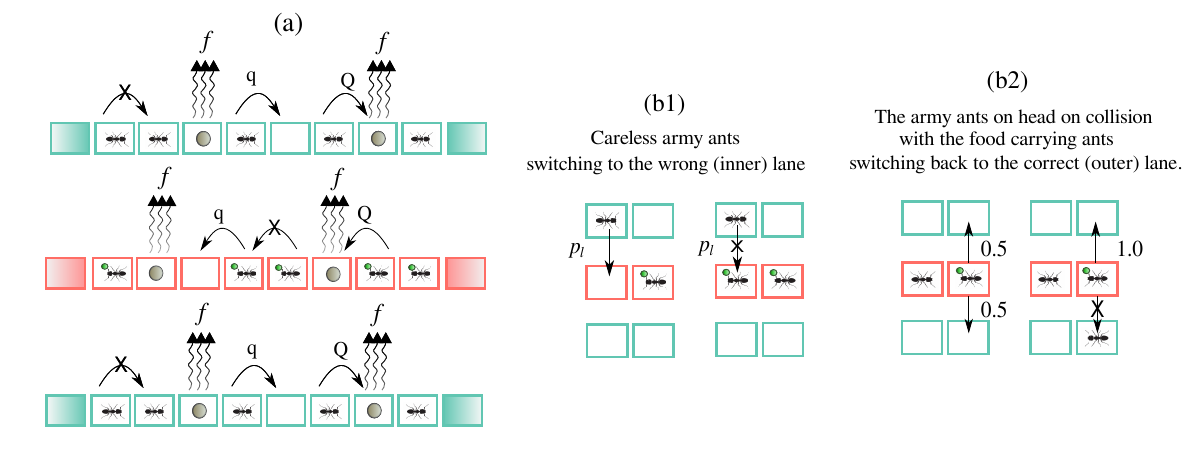} 
\end{center} 
\caption[scale=tiny]{(Color online) (a) Schematic depiction of the kinetic processes and the corresponding rates  
associated with the movement of (i) army ants from left to right along the outer lanes (green in color), and  
(ii) food carrying army ants from right to left along the inner lane (red in color). An ant can move to an empty 
site ahead of it with (i) probability \textit{Q} in the presence of pheromone, and (ii)  probability $q$ in the 
absence of pheromone, at the target site. Moreover, pheromone existing at an empty site evaporates with 
probability $f$. 
The lane changing rules are depicted schematically in (b1) and (b2). As depicted in (b1), an army ant can move
from an outer lane to the inner (wrong) lane with probability $p_{\ell}$ if the adjacent site in inner lane is empty. 
As shown in (b2), if an army ant that erroneously moves to the inner lane collides head-on with an
oncoming food carrying ant, then the defaulter army ant can move out to adjacent sites in either of the outer lanes 
with equal probability if both the adjacent sites of outer lanes are empty or with certainty to the only available empty 
site on the adjacent outer lane. (Fig.4 of \cite{pradhan21})} 
\label{fig-Bidor3LaneModel2}
\end{figure}

The mechanisms of the emergence of multi-lane bi-directional traffic of ants were reported by several research groups; for example, army ants {\it Eciton burchelli} was shown to self-organize into a 3-lane traffic (see Fig.\ref{fig-Bidor3LaneModel1}) \cite{couzin02}. Motivated by the curiosity as to the possible effects lane changing and overtaking slower ants on congested lanes in such 3-lane traffic ASEP-based theoretical models have been developed \cite{pradhan21} (see Figs.\ref{fig-Bidor3LaneModel1} and \ref{fig-Bidor3LaneModel2} for the details). Some of the unusual features of the fundamental diagram were subsequently explained by further deeper analysis.

\section{Human pedestrian traffic: Floor-field model inspired by stigmergy in ant traffic} 

We have already mentioned the similarities and differences between collective transport of a single vesicular cargo by multiple molecular motors and that of a single piece of leaf or food by multiple ants. We have also drawn attention to the similarities and differences between the traffic-like movement of molecular motors on filamentous tracks and that of ants on a trail. In the same spirit, in this section, we discuss how the some analogies between the traffic of ants and humans can be exploited for modeling the latter. It is because of this analogy that we discuss human pedestrian traffic in the context of physics of life. However, it is worth pointing out that exploiting analogy between the ant societies and human societies is not limited to modeling only traffic-like phenomena. How human societies rise, thrive and fall has also been analyzed by drawing analogies between ant swarm \cite{moffettAntSwarmbook} and `human swarm' \cite{moffettHumanSwarmbook}.  The broader subject of human swarm \cite{moffettHumanSwarmbook} and human mobility \cite{barbosa18} may have stronger overlap with cognitive, behavioral and social sciences \cite{feliciani23,moffettHumanSwarmbook} and is, therefore, beyond the scope of this brief article.

 Historically, models of human pedestrian traffic have been developed often drawing analogy with those of vehicular traffic \cite{schadschneider19,corbetta23}. Corbetta and Toschi \cite{corbetta23} have classified the theoretical models of the dynamics of human crowds into those at five different scales, depending on the spatial and temporal resolution as well as the detailed description of the state of the system. Out of these there are two main approaches both of which are `microscopic', i.e., the movement of each individual is described explicitly. The first approach is similar to classical Newtonian mechanics where the dynamics of each individual is described by a 2nd order ordinary differential equation. Pedestrians interact with each other and with their environment through forces. Not all of these forces is necessarily physical interaction; the so-called ``social forces'' capture the normal human desire to avoid too close contact with other persons in a crowd. We will not discuss this approach here; interested readers are referred to the ref.\cite{chen18}.  The other popular class of `microscopic' models are based on cellular automata. In this class, the floor field model (FFM) \cite{burstedde01,kirchner02} has become paradigmatic. This model was inspired by {\it stigmergy} \cite{theraulaz99,heylighen16a,heylighen16b}, the mechanism of indirect communication between ants using the pheromones.

In a continuum model the movement of individual ants would be described by a classical Langevin equation where the gradient of the pheromone field 
$\sigma({\vec r},t)$ at a location $\vec{r}$ will capture the pheromone-induced bias on the movement of the ant at that location. Since pheromone-field is created, and reinforced, by the ants the dynamical equation for the pheromone field would be coupled to that of the ants. Moreover, the pheromones evaporate and diffuse; so, the dynamical evolution of the pheromone field would be governed by a reaction-diffusion type partial differential equation (see ref.\cite{kebook} for a general introduction). 

\begin{figure}[t] 
\begin{center}
\includegraphics[width=0.65\columnwidth]{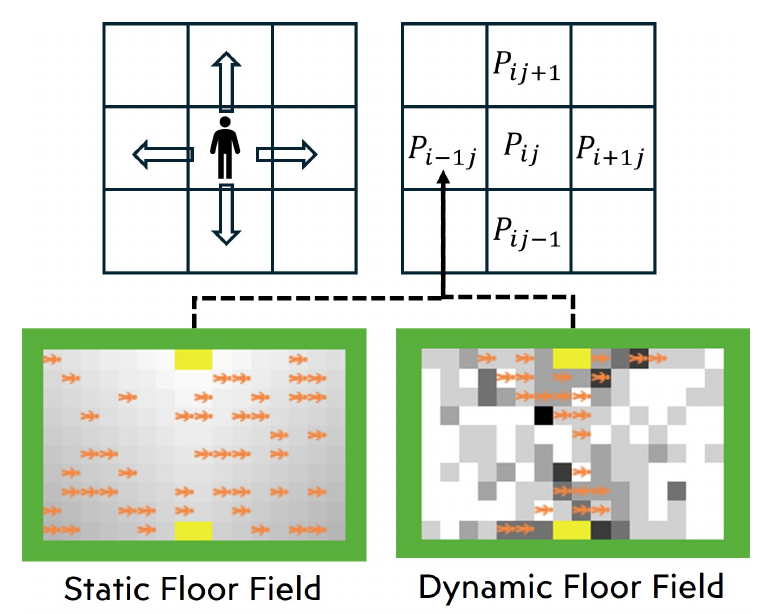} 
\end{center} 
\caption[scale=tiny]{(Color online) Schematic description of the Floor-field model (see the text for details).} 
\label{fig-FFM}
\end{figure}

The same ideas are implemented in a more simplified manner in the floor-field model \cite{burstedde01,kirchner02} for human pedestrian dynamics.
In this model, a virtual pheromone field is created by moving pedestrians on a floor that is assumed to be discrete. The static floor field $S_{ij}$ does not change with time; it encodes information about the environment and the desired direction of motion. In contrast, the dynamic floor field, denoted by the symbol $D_{ij}$, is time-dependent. Besides being created fresh by pedestrians moving into the position $i,j$, the virtual pheromone at $i,j$ also diffuses and decay, tending to result in its dilution and, even disappearance; this dynamics is governed by the updating of the dynamic floor field $D_{ij}$  \cite{scnbook,burstedde01,schadschneider19,nishinari04}

The relative influence of the two floor fields is controlled by coupling constants $k_S$ and $k_D$. Time is also treated as a discrete variable. The dynamics of the pedestrians is formulated in terms of update rules using the language of cellular automata where the probabilities of transition from one state of the system to another depend on the local configuration of the pedestrians (see Fig.\ref{fig-FFM}). The instantaneous transition probabilities depend on the state of the dynamic floor field. The generic form of the transition probability $p_{ij}$ to a neighbouring cell $(i,j)$ is given by
\begin{equation}
    p_{ij} = N e^{k_S S_{ij}}  e^{k_D D_{ij}} (1-n_{ij}) \xi_{ij}\,.
\label{eq-FFM}
\end{equation}
where $n_{ij}= 0,1$ is the number of persons in cell $(i,j)$. The factor  $(1 - n_{ij})$ in Eq.(\ref{eq-FFM}) ensures that no transition to an occupied cell takes place.
The symbol $\xi_{ij}$ accounts for the walls or obstacles: $\xi_{ij}=0$ for cells $i,j$ whose occupation is forbidden (like walls or obstacles), whereas $\xi_{ij}=1$ otherwise. $N$ is a normalization constant that ensures the sum of all transition probabilities for a pedestrian is 1. If $k_S \gg k_D$, the static floor field dominates and  the pedestrians choose the shortest path to the exit. On the other hand, if $k_D \gg k_S$, the dynamic floor field dominates and in that case the pedestrians tend to follow other moving pedestrians; this is known as herding behaviour. \\

\section{Conclusion} 

In this brief article we have summarized some of the major directions of research in the area of collective transport and traffic-like phenomena in biology. The systems covered span a wide range of length scales starting from traffic of nano-motors to human pedestrians. The similarities and differences between the phenomena at different scales have been pointed out and common themes have been emphasized. Because of the constraint on the length of the article, selection of the works reviewed here was somewhat biased in favour of our own contributions over the last 20 years. But, major contributions of other research groups have also been cited. We hope the next 20 years will see serious efforts in testing many of the novel theoretical predictions and further exchange of ideas across disciplinary boundaries will enrich our understanding of physics of life.

{\bf Acknowledgements}:

Over the last ten-years of the 20 year period since the publication of ref.\cite{csk05polrev} in Physics of Life Reviews, SERB, Government of India, has generously supported research of DC at IIT Kanpur through a J.C. Bose National Fellowship awarded to him. DC also thanks the Visitors Program of the Max-Planck Institute for Physics of Complex Systems for hospitality during many research visits to Dresden, Japan Society for the Promotion of Science for Invitational Fellowship during Visiting Professorship at the University of Tokyo and the International Centre for Theoretical Sciences of TIFR for hospitality in Bangalore during several ICTS programs. We acknowledge the contributions of all our research collaborators, students and postdocs who co-authored our papers reviewed or cited in this article. 
We apologize to those authors some of whose works in the last 20 years, though no less important, could not be reviewed here because of constraint on the length of this article.


\end{document}